\documentclass[fleqn,usenatbib]{mnras}

\usepackage{newtxtext,newtxmath}

\usepackage[T1]{fontenc}

\DeclareRobustCommand{\VAN}[3]{#2}
\let\VANthebibliography\thebibliography
\def\thebibliography{\DeclareRobustCommand{\VAN}[3]{##3}\VANthebibliography}

\usepackage{graphicx}	
\usepackage{amsmath}	
\usepackage{flushend}

\title[SGWB constraints from Gaia DR3]{Stochastic gravitational wave background constraints from Gaia DR3 astrometry}

\author[S. Jaraba, J. Garc\'ia-Bellido, S. Kuroyanagi, S. Ferraiuolo, M. Braglia]{
Santiago Jaraba,$^{1}$\thanks{E-mail: santiago.jaraba@uam.es}
Juan Garc\'ia-Bellido,$^{1}$\thanks{E-mail: juan.garciabellido@uam.es}
Sachiko Kuroyanagi,$^{1, 2}$\thanks{E-mail: sachiko.kuroyanagi@csic.es}
Sarah Ferraiuolo,$^{3}$\thanks{E-mail: sarah.ferraiuolo@studio.unibo.it}
Matteo Braglia$^{4}$\thanks{E-mail: mb9289@nyu.edu}
\\
$^{1}$Instituto de F\'isica Te\'orica UAM-CSIC, Universidad Aut\'onoma de Madrid, Cantoblanco 28049 Madrid, Spain\\
$^{2}$Department of Physics and Astrophysics, Nagoya University, Nagoya, 464-8602, Japan\\
$^{3}$Dipartimento di Fisica e Astronomia, Alma Mater Studiorum, Universit\`a di Bologna, Via Gobetti, 93/2, I-40129 Bologna, Italy\\
$^{4}$Center for Cosmology and Particle Physics, New York University, 726 Broadway, New York, NY 10003, USA
}

\date{Accepted XXX. Received YYY; in original form ZZZ}

\pubyear{2023}

\begin{document}
\label{firstpage}
\pagerange{\pageref{firstpage}--\pageref{lastpage}}
\maketitle

\begin{abstract}
Astrometric surveys can be used to constrain the stochastic gravitational wave background (SGWB) at very low frequencies. We use proper motion data provided by Gaia DR3 to fit a generic dipole+quadrupole field. We analyse several quasar-based datasets and discuss their purity and idoneity to set constraints on gravitational waves. For the cleanest dataset, we derive an upper bound on the (frequency-integrated) energy density of the SGWB $h_{70}^2\Omega_{\rm GW}\lesssim 0.087$ for $4.2\times 10^{-18}~\mathrm{Hz}\lesssim f\lesssim 1.1\times 10^{-8}~\mathrm{Hz}$. We also reanalyse previous VLBI-based data to set the constraint $h_{70}^2\Omega_{\rm GW}\lesssim 0.024$ for $5.8\times 10^{-18}~\mathrm{Hz}\lesssim f\lesssim 1.4\times 10^{-9}~\mathrm{Hz}$ under the same formalism, standing as the best astrometric constraint on GWs. Based on our results, we discuss the potential of future Gaia data releases to impose tighter constraints.
\end{abstract}




\section{Introduction}
\label{sec:introduction}

Stochastic gravitational wave backgrounds (SGWBs) have received increasing attention as potential probes of the early universe. There have been many attempts to place constraints on the amplitude of the SGWB at different frequencies. For example, CMB B-mode measurements have placed an upper limit on the tensor-to-scalar ratio $r<0.036$ (95\% CL) at $k=0.05~{\rm Mpc}^{-1}$~\citep{BICEP:2021xfz}, which corresponds to $\Omega_{\rm GW}\lesssim 1.5 \times 10^{-16}$ at $f = 7.7 \times 10^{-17}$ Hz. Advanced LIGO’s and Advanced Virgo’s third observing run (O3) combined with upper limits from the earlier O1 and O2 runs has provided $\Omega_{\rm GW}\lesssim 5.8 \times 10^{-9}$ (95\% CL) for a flat spectrum in the frequency band $20-76.6$ Hz~\citep{KAGRA:2021kbb}. Furthermore, pulsar timing arrays indicated a signal which is compatible with an $\Omega_{\rm GW}\sim10^{-9}$ at $f=1\,{\rm yr}^{-1}$, although there is no statistical evidence for the SGWB hypotesis~\citep{NANOGrav:2020bcs,Chen:2021rqp,Goncharov:2021oub,Antoniadis:2022pcn}. Future methodologies that have been proposed for testing a SGWB include space-based interferometers such as LISA~\citep{LISACosmologyWorkingGroup:2022jok}, using the Galactic Bulge Time Domain (GBTD) survey of the Nancy Grace Roman Space Telescope~\citep{Wang:2022sxn} and using resonances in binary systems~\citep{Blas:2021mpc}.

Gravitational wave astrometry is another way to probe low frequency SGWBs. Precise astrometric measurements, such as those conducted by the radio-based interferometers and Gaia mission, offer a novel approach for detecting SGWB. This is possible because GWs cause fluctuations in the positions and proper motions of stars with a characteristic pattern in the sky~\citep{Linder:1986fdo,Braginsky:1989pv,Fakir:1993bj,Flanagan:1993ix,Pyne:1995iy,Kaiser:1996wk,Damour:1998jm,Jaffe:2004it,Schutz:2010lmv,Book:2010pf, Qin:2018yhy}. Some previous works have succeeded in getting an upper bound on the SGWB amplitude using astrometry. \cite{Gwinn:1996gv} analysed limits on quasar proper motions obtained from VLBI astrometric data, reporting $\Omega_{\rm GW}\lesssim 10^{-1}$ for $10^{-17}~\mathrm{Hz}\lesssim f\lesssim 10^{-9}~\mathrm{Hz}$. \cite{Titov:2010zn} updated this limit to $\Omega_{\rm GW}\lesssim 10^{-2}$. \cite{Darling:2018hmc} used the VLBA (Very Long Baseline Array) data in~\cite{Truebenbach:2017nhp} to get a 95\% upper bound of $\Omega_{\rm GW}\lesssim 0.0064$ for the frequency range $6\times 10^{-18}~{\rm Hz}\lesssim f\lesssim 10^{-9}~{\rm Hz}$. More recently, an upper bound $\Omega_{\rm GW}\lesssim 4.8\times 10^{-4}$ for $f\lesssim 10^{-9}~{\rm Hz}$ was reported by~\cite{Aoyama:2021xhj}. In a similar frequency band, \cite{Linder1988} set an upper bound $\Omega_{\rm GW}\lesssim 10^{-3}$ for $10^{-16}~\mathrm{Hz}\lesssim f\lesssim 10^{-13}~\mathrm{Hz}$ using galaxy correlations.

The European Space Agency’s (ESA) \textit{Gaia} mission~\citep{Gaia:2016zol} is a promising experiment for the purpose of setting astrometric constraints on the SGWB. It was launched in December 2013 and is expected to operate until 2025. In December 2020, the Gaia collaboration published their Early Data Release 3 (EDR3)~\citep{GaiaEDR3}, containing data from around 1.8 billion objects based on 34 months (1038 days) of operations. This data release includes full astrometric data for 1.468 billion sources~\citep{GaiaEDR3astrom}. More recently, in June 2022, the Gaia Data Release 3 (DR3)~\citep{Gaia:2022,Gaia:2022val} was published, expanding the information of Gaia EDR3 sources and including classification of objects. Application of GW astrometry to Gaia data has been extensively discussed in the literature~\citep{Moore:2017ity,Klioner:2017asb,Mihaylov:2018uqm,Mihaylov:2019lft}.

In this paper, we aim to derive an upper bound on the SGWB amplitude by using the latest release by Gaia. To do so, we combine the Gaia EDR3 proper motion data and the source information from DR3 to produce datasets based on quasars, which are expected to have small intrinsic proper motion. We then build a likelihood to fit the proper motion to a generic dipole+quadrupole pattern to each dataset we consider, and sample it via Markov Chain Monte Carlo (MCMC) to extract constraints on the multipole coefficients. From the posterior distribution of the total power from all the quadrupole coefficients, we derive a 95\% upper bound which we then convert into SGWB amplitude, addressing its frequency range validity. The results of this paper provide the most updated astrometric constraints on the amplitude of the SGWB. In addition, we revisit the datasets used in previous works~\citep{Darling:2018hmc,Aoyama:2021xhj} and update earlier upper bounds.

This paper is structured as follows. In Sec.~\ref{sec:methodology}, we review  the method to place an upper bound on the SGWB amplitude using proper motion of distant sources. In Sec.~\ref{sec:datasets}, we describe the different datasets we work with. In Sec.~\ref{sec:results}, we present the results and compare them to the upper bounds set by previous works. We conclude with a discussion on the prospects for future Gaia data releases and other proposed surveys such as Theia~\citep{Theia:2017xtk, Malbet:2022lll, Garcia-Bellido:2021zgu}. For reference, in Appendix~\ref{app:triangle_plots}, we collect all the triangle plots from our MCMC analysis, as well as the estimated covariance matrices for the parameter correlations.

\section{Methodology}
\label{sec:methodology}

Here we summarize the formalism on how to obtain a SGWB upper bound from astrometric measurements. A good review is~\cite{Book:2010pf}.

\subsection{SGWB from astrometry}
\label{ssec:sgwb_astrometry}

When an observer looks at a distant star, the observed photons are the result of their history from the time they were emitted by the source. As light travels to us, the spacetime in which it propagates is subject to the laws of General Relativity. This implies that the passage of a gravitational wave can bend the trajectory of a light ray, generating an angular deflection in the position at which we measure this light ray from Earth. This would be impossible to measure by monitoring a single source, but it is theoretically possible if we do it with a large number of sources and test the angular correlations between different sources.

Here we are not interested in the angular deflection that a single gravitational wave produces, but in the one induced by a SGWB coming from all directions. For this SGWB to leave an imprint on observed sources, it must have a low frequency ($\lesssim$nHz). According to~\cite{Book:2010pf}, we can expect a correlated angular deflection
\begin{equation}
    \label{eq:delta_n}
	\langle\delta\vec{n}^2\rangle=\frac{H_0^2}{4\pi^2}\int d\ln f\,\frac{\Omega_{\rm GW}(f)}{f^2},
\end{equation}
where $\vec{n}$ is the position vector of each source and $\delta\vec{n}$ the angular deflection produced by gravitational waves. This expression assumes that the distance to the source is much greater than the wavelength of the gravitational wave, in the so-called the distant source limit.

By taking the time derivative, we can obtain an equivalent equation for the proper motion,
\begin{equation}
	\label{eq:pm_spectrum}
	\langle\delta\dot{\vec{n}}^2\rangle=H_0^2\int d\ln f\Omega_{\rm GW}(f).
\end{equation}

If we assume that we have data from positions measured over a time period $T\sim 1/f$, we can estimate
\begin{equation}
    \label{eq:OmegaGW_pm}
    \Omega_{\rm GW}(f)\sim\langle\delta\dot{\vec{n}}^2(f)\rangle/H_0^2,
\end{equation}
or at least provide an upper bound when our sample is dominated by systematic uncertainties. The frequency range validity of this bound is discussed in the next section.

We should also mention that Eq.~\eqref{eq:OmegaGW_pm} has been used in the literature to set rough estimates for the expected SGWB constraints from astrometry within a certain mission. In~\cite{Book:2010pf}, it is assumed that, for a dataset of size $N$ with position uncertainties $\Delta\theta$, we can expect to detect a root mean square angular speed of $\Delta\theta/(T\sqrt{N})$ and therefore estimate an upper bound as 
\begin{equation}
	\label{eq:omegagw_bound}
	\Omega_{\rm GW}\lesssim\frac{\Delta\theta^2}{NT^2H_0^2}.
\end{equation}

\subsection{Frequency range validity}
\label{ssec:frequency_range}

The frequency of GWs probed by this method is determined by the observation time $T$. When we have only the proper motions averaged during the observation time, which is the case of Gaia DR3, we do not see the contribution from the higher frequencies ($f \gtrsim 1/T$) as rapidly oscillating effects are canceled out. On the other hand, for lower frequency GWs ($f \lesssim 1/T$), although we cannot track the oscillation of the wave within the observation time, the linear time evolution in the tensor metric perturbations with a quadrupole pattern still induce spatially correlated fluctuations in the proper motions.
From this, we can constrain the total contribution from all the low frequencies $\int_{f\lesssim 1/T} d\ln f ~ \Omega_{\rm GW}(f)$~\citep{Book:2010pf}. Note that Eq.~\eqref{eq:pm_spectrum} only holds in the distance source limit, which imposes that the distance to each source must be much greater than the GW wavelength. Thus, the lowest frequency contributing to our measurement is $f\sim c/D_{\rm min}$ with $D_{\rm min}$ being the minimum distance among all the sources. In~\cite{Darling:2018hmc}, this assumption is relaxed, assuming that it is enough if the wavelength is below the distance to 75\% of the sources, i.e., the maximum wavelength is the first quartile of the distance distribution. We also make this assumption, which is further discussed in Sec.~\ref{ssec:redshifts}.

The situation will change when the full Gaia data is released in DR5, and we will have access to each source position time series. In that case, our minimum time resolution is given by the cadence of our time series, $\Delta t$, which extends the higher frequency bound as $f\lesssim 1/\Delta t$. In addition, having time series, we see the oscillation in time and no longer be interested in a linear change in time. Thus, one may set the lower frequency bound as $f\gtrsim 1/T$, as seen in the sensitivity curves provided in the references addressing astrometric time series, e.g.,~\cite{Moore:2017ity,Wang:2022sxn, Klioner:2017asb}.

\subsection{Vector spherical harmonics}
\label{ssec:vsh}

In order to characterize the proper motion field, we follow a very similar procedure to the one in~\cite{Darling:2018hmc}. Our dataset consists of International Celestial Reference System (ICRS) coordinates (right ascension $\alpha$ and declination $\delta$) of sources and their proper motions and errors within this system.

First, we need to decompose our proper motion field into spherical harmonics. We use the vector spherical harmonics definition in~\cite{Mignard:2012xm}, which divides the basis into \textit{spheroidal} $\vec{S}_{lm}$ (or \textit{electric}) and \textit{toroidal} $\vec{T}_{lm}$ (or \textit{magnetic}) modes
\begin{align}
    \vec{S}_{lm}(\alpha,\delta)&=\frac{1}{l(l+1)}\nabla Y_{lm}(\alpha, \delta),\\
    \vec{T}_{lm}(\alpha,\delta)&=-\frac{1}{l(l+1)}\hat{n}\times\nabla Y_{lm}(\alpha, \delta).
\end{align}

Any complex-valued vector field can be expressed as a series of these basis vectors with complex multipole coefficients $s_{lm}$, $t_{lm}$, $l\geq 1$, $|m|\leq l$. In general, when refering to either of these modes, we use $\vec{R}_{lm}$ for the spherical harmonics basis and $r_{lm}$ for the multipole coefficients. In particular, the constraints for a real-valued vector field allow to write
\begin{equation}
	\label{eq:multipoles}
    \vec{V}(\alpha,\delta) =\!\! \sum_{\substack{r=s,t\\R=S,T}}\!\sum_{l=1}^\infty\left[r_{l0}\vec{R}_{l0} + 2\sum_{m=1}^l\left(r_{lm}^{\rm Re}\vec{R}_{lm}^{\rm Re}-r_{lm}^{\rm Im}\vec{R}_{lm}^{\rm Im}\right)\right].
\end{equation}

Given that the main contribution to the SGWB is given by the quadrupole (5/6 of the total amplitude~\citep{Pyne:1995iy,Book:2010pf}), we use only this multipole to compute the upper bound on $\Omega_{\rm GW}$. From Eq.~\eqref{eq:OmegaGW_pm}, we can relate $\Omega_{\rm GW}$ to the quadrupole power as~\citep{Darling:2018hmc}
\begin{equation}
	\label{eq:quadrupole_to_omega}
    \Omega_{\rm GW}=\frac{6}{5}\frac{1}{4\pi}\frac{P_2}{H_0^2}=0.000438\frac{P_2}{(1~\mathrm{\mu as/yr})^2}h_{70}^{-2},
\end{equation}
with $P_2$ the total power in the $l=2$ multipole,
\begin{align}
    \label{eq:power}
    P_l=&P_l^s + P_l^t,\\
    \label{eq:power2}
    P_l^r=\sum_{m=-l}^l|r_{lm}|^2=r_{l0}^2&+2\sum_{m=1}^l(r_{lm}^{\rm Re})^2+(r_{lm}^{\rm Im})^2,
\end{align}
where $r$ stands for either the $s$ or $t$ coefficients. For a quadrupole pattern from a SGWB, we expect both modes to contribute equally~\citep{Book:2010pf}.

We have also used $H_0=70h_{70}~\mathrm{km~s^{-1}~Mpc^{-1}}=14.76h_{70}~\mathrm{\mu as/yr}$. Note that the 0.000438 factor slightly differs from the 0.00042 in~\cite{Darling:2018hmc} because they use $H_0\approx 15h_{70}~\mathrm{\mu as/yr}$.

We then aim to compute the $l=2$ coefficients in Eq.~\eqref{eq:multipoles}, which can be obtained by fitting the data to a generic quadrupole. Another possible approach if the dataset has good sky coverage would be to divide the sky in a certain number of cells, e.g., by using HEALPix (Hierarchical Equal Area isoLatitude Pixelization)\footnote{\url{https://healpix.sourceforge.io}} or similar methods, like the one suggested in~\cite{Moore:2017ity}. In this case, each cell is assigned the average proper motion of the sources in that cell, thus compressing the data. Then, the quadrupole can be computed by using the orthogonality of spherical harmonics. However, this method assigns the same weight to all cells, regardless of the number of sources they have, and the effects of empty cells must be addressed.

In our case, we have $N\sim10^6$ and the sky coverage is good enough to use the latter method. However, for all our datasets, we have checked the HEALPix method to provide multipole coefficients with amplitudes around a factor 10 higher than for the ones obtained from fitting. We then choose the fitting method due to its higher accuracy, following existing literature~\citep{Truebenbach:2017nhp,Darling:2018hmc,Aoyama:2021xhj}.

\subsection{Fitting the data}
\label{ssec:fitting}

For a given dataset, we fit both the dipole and quadrupole coefficients simultaneously. For fitting data, we use the Python package \textit{emcee}~\citep{Foreman-Mackey:2012any}, which performs an MCMC over the data. We set very generous uniform priors in the $[-100, 100]~\mathrm{\mu as}$ range for all coefficients of the vector spherical harmonics and use $10\times n$ walkers, with $n=16$ being the number of parameters. In order to test convergence, we compute the autocorrelation times $\tau_i$~\citep{Goodman2010}, with $i$ denoting parameter index, and run the MCMC for at least $100\tau$ iterations, with $\tau=\max_i\{\tau_i\}$. In practice, this amounts to around 20,000 iterations for the dipole+quadrupole fits. We then burn the first $2\tau$ iterations in order to obtain our posterior distribution.

As addressed in~\cite{Darling:2018hmc}, any astrometric dataset shows significant, uncorrelated proper motions, which makes it challenging to obtain the underlying correlated proper motion background with a much smaller amplitude. It is therefore necessary to use a permissive method that does not get carried away by such uncorrelated noises. By assuming that the estimated experimental uncertainties are only valid as lower bounds of the real uncertainties, one can get the likelihood proposed in~\citep{Sivia:2006} and used in~\cite{Darling:2018hmc},
\begin{equation}
    \label{eq:logl}
	\ln\mathcal{L} = \text{const.} + \sum_{i=1}^N\ln\left(\frac{1-e^{-R_i^2/2}}{R_ i^2}\right),
\end{equation}
where $R_i$ is the standard residual $R_i=(D_i-\text{Model})/\sigma_i$ with $D_i$ being a data point $D_i$ and $\sigma_i$ its error. By following \cite{Mignard:2012xm}, we define the residual as 
\begin{equation}
	R_i^2=\left(\frac{D_i^\alpha-V_\alpha}{\sigma_i^\alpha}\right)^2 + \left(\frac{D_i^\delta-V_\delta}{\sigma_i^\delta}\right)^2,
\end{equation}
with $\vec{V}$ being the vector defined in Eq.~\eqref{eq:multipoles} truncated to a range $l_{\rm min}\leq l\leq l_{\rm max}$. 

While our goal is to fit the quadrupole, there may be a residual dipole component that may impact our results if we fit the quadrupole directly. In our results, this dipole has a power $\sqrt{P_1/P_2}\sim 1-4$, which is non-negligible (see Table~\ref{tab:multipoles}). In~\cite{Darling:2018hmc}, a first MCMC fits a generic dipole alone, which is then subtracted from the proper motion data. Next, a second MCMC fits the quadrupole to the dipole-subtracted data. In this paper, we take another approach, which is simultaneously fitting the dipole and quadrupole. This would account for possible correlations between both multipoles, even if they are small, producing a more accurate fit. We have checked this method to worsen the constraints by around 10\%, as expected from the larger parameter space.

\subsection{Statistical significance}
\label{ssec:statistical_significance}

The statistical significance of the fitted quadrupole is tested both from the frequentist (Z score) and Bayesian (Bayes factor) approaches.

\subsubsection{The Z score}
\label{sssec:z_score}

The key quantity to compute is the quadrupole power defined in Eq.~\eqref{eq:power}, so it is essential to understand its statistical behavior and how it varies when the dataset is pure noise and when it is not.

If the dataset is pure isotropic Gaussian noise, the multipole coefficient posterior distributions can be assumed to follow independent Gaussian distributions with a mean around zero and a certain variance. Ideally, for a given multipole, if ${\rm Var}(r_{l0})=\sigma_{l0}^2$, then ${\rm Var}(r_{lm}^{\rm Re})\approx{\rm Var}(r_{lm}^{\rm Im})\approx\sigma_{lm}^2/2$~\citep{Mignard:2012xm}, which can also be observed in our posteriors, even when they are not centered near zero (see Table~\ref{tab:multipoles}).

As a result, when we compute the power from Eq.~\eqref{eq:power2}, the factors 2 make the operation \textit{a sum of squares of zero mean, same variance, independent Gaussian distributions}. In other words, if we assume pure isotropic Gaussian noise, the power of a given multipole should follow a \textit{chi-squared distribution} $\chi^2_n$ with $n=2(2l+1)$ degrees of freedom, rescaled by $\sigma_0$ like $f_{P_l}(x)=f_{\chi^2_k}(x/\sigma_0)/\sigma_0$, where $f$ stands for the probability distribution function.

\begin{figure}
    \centering
    \includegraphics[width=\linewidth]{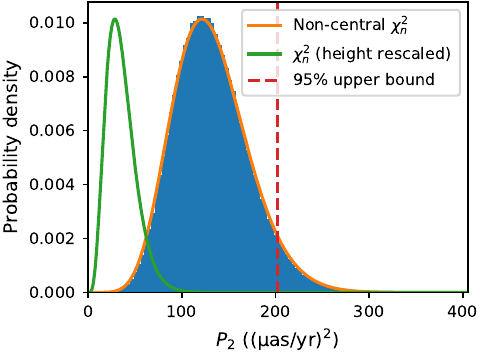}
    \caption{Posterior distribution for the quadrupole power $P_2$ for the astrometric dataset (see Sec.~\ref{sec:datasets} below), with the corresponding chi-square and non-centered chi-square distributions superimposed.}
    \label{fig:P2_posteriors}
\end{figure}

If, however, our posteriors are not fitting pure isotropic random noise, what we get is \textit{a sum of squares of non-zero mean, same variance, independent Gaussian distributions}, which produces a \textit{non-central chi-squared distribution} with the same degrees of freedom and rescaling, and with the non-centrality parameter being the power computed from the $r_{lm}$ means. One can see the difference in Fig.~\ref{fig:P2_posteriors}, where we plot the quadrupole power posteriors for one of our datasets, together with the usual and non-central chi-squared distributions and indicating the 95\% upper bound we use to set our constraint in $\Omega_{\rm GW}$, from Eq.~\eqref{eq:quadrupole_to_omega}.

In order to test how likely it is to obtain our maximum likelihood estimate (MLE) for $P_l$ assuming pure Gaussian noise, one can first compute the quantity
\begin{equation}
    W_l=\left(\frac{r_{l0}}{\sigma_{l0}}\right)^2+\sum_{m=1}^l\left[\left(\frac{r_{lm}^{\rm Re}}{\sigma_{lm}^{\rm Re}}\right)^2+\left(\frac{r_{lm}^{\rm Im}}{\sigma_{lm}^{\rm Im}}\right)^2\right],
\end{equation}
where the $r_{lm}$ are the MLE. This equation is similar to Eq.~\eqref{eq:power2}, only with each coefficient variance rescaled to one. Under the assumption of pure white noise, this quantity should behave as a chi-square with no rescaling, so one could simply compute $P_{\chi^2_n}(X>W_l)$ and see how unlikely it is to obtain the computed $W_l$. Instead, one can discuss in terms of sigmas of a Gaussian distribution, which is more common in Physics. For this purpose, one should realise that~\citep{Wilson:1931}
\begin{equation}
    \left(\frac{\chi^2_n}{n}\right)^{1/3}\sim N\left(1-\frac{2}{9n},\sqrt{\frac{2}{9n}}\right)
\end{equation}
to a good degree of approximation increasing with $n$, where $N(\mu,\sigma)$ denotes the Gaussian distribution of mean $\mu$ and variance $\sigma^2$. Therefore, the Z score,
\begin{equation}
    Z=\sqrt{\frac{9n}{2}}\left[\left(\frac{W_l}{n}\right)^{1/3}-\left(1-\frac{2}{9n}\right)\right],
\end{equation}
is a good measurement of how many sigma we are away from the pure noise hypothesis. This statistic was defined in~\cite{Mignard:2012xm} and is used in~\cite{Darling:2018hmc}, so we also compute it for our fits.

\subsubsection{The Bayes factor}
\label{sssec:bayes_factor}

The Z score for $l=2$ is the frequentist approach to addressing statistical significance of the fitted quadrupole. However, we run MCMCs within a Bayesian framework, so it is thus relevant to provide also a Bayesian estimator for the statistical significance of our results.

After the MCMC runs, the posterior distribution for the multipole coefficients approximately follows a multivariate Gaussian distribution. From its covariance matrix $\Sigma_{12}$ and the maximum likelihood $\mathcal{L}_{12}$, we can compute Bayes factors using the approach described in~\cite{Garcia-Bellido:2023yqk}.

We aim to test whether the fitted quadrupole is statistically relevant. Therefore, our alternative hypothesis will be that the dataset contains a dipole and quadrupole (full fit) and the null hypothesis will be that there is only dipole. To compute the evidence for null hypothesis, we do not perform a separate fit to the dipole. Instead, we assume the correlations between the dipole and quadrupole to be negligible, so the likelihood $\mathcal{L}_1$ computed from the dipole of our MLE and the $(l=1,l=1)$ submatrix $\Sigma_1\subset\Sigma_{12}$ will suffice. For our four main datasets and also for the VLBA and VLBA+Gaia DR1, the low correlation can be seen in Fig.~\ref{fig:cov_matrices} in the Appendix.

We also assume that the prior range $[-L,L]$ ($L=100~\mathrm{\mu as/yr}$) is wide enough to neglect the posterior probability outside this region for all the multipole coefficients, which is reasonable given our posteriors (see Figs.~\ref{fig:posteriors_ours}-\ref{fig:posteriors_SDSS}). Under these assumptions, one can integrate the multivariate Gaussians to compute the evidence of each hypothesis and show that

\begin{align}
    \mathcal{B}^{12}_1\approx\frac{\mathcal{L}_{12}}{\mathcal{L}_1}\left(\frac{\sqrt{2\pi}}{2L}\right)^{n_{12}-n_1}\left(\frac{\det(\Sigma_{12})}{\det(\Sigma_1)}\right)^{1/2},
\end{align}
where $n_{12}=16$ and $n_1=6$ are the degrees of freedom of the models.

\section{The datasets}
\label{sec:datasets}

We use quasar-based datasets from the Gaia DR3, since they have small intrinsic proper motions compared to the rest of the sources. Even if the full Gaia DR3 has more than a billion sources, the drawback of using data with high intrinsic proper motions overcomes the advantage of having such a large dataset. Note that the number of sources $N$ enters Eq.~\eqref{eq:omegagw_bound} just as $N^{-1}$, whereas $\Delta\theta$ appears as $(\Delta\theta)^2$. For reference, the averaged proper motion coordinates of the full Gaia catalog (O($10^9$) sources) are of order 1 mas/yr, while we reduce them to order 1 $\mu$as/yr for one of our datasets with O($10^6$) sources. According to Eq.~\eqref{eq:omegagw_bound}, this would imply a better constraint by a factor 1000.

We note that the Gaia collaboration has not provided any official Quasi Stellar Object (QSO) catalogs, but only a subsample of the Gaia DR3 dataset labeled as \textit{QSO candidates}. The latter comprises a compilation of sources considered as quasars by different modules. In order to build our own dataset, we follow the selection criteria detailed in~\cite{Gaia:2022vcs} and in the Gaia DR3 documentation website\footnote{Gaia DR3 online documentation: \url{https://www.cosmos.esa.int/web/gaia-users/archive/gdr3-documentation}}, which we summarize below:
\begin{itemize}
    \item The main contribution to this catalog is provided by Gaia Discrete Source Classifier (DSC)~\citep{Delchambre:2022ugo}, which classifies sources into five classes: quasar, galaxy, star, white dwarf, and physical binary star. DSC consists of three classifiers: Specmod, which uses BP/RP spectra; Allosmod, which uses other features such as parallax, proper motions, or colour indices, and Combmod, which combines the output class probabilities of both of them and assigns a combined probability, labeled as \texttt{classprob\_dsc\_combmod\_quasar}. We denote this combined probability as $p_{\rm QSO}$ in the following sections.
    
    A source enters the QSO candidates list if any of the three classifiers assign a probability of being a quasar above 0.5. In addition, the Gaia Quasar Classifier module (QSOC)~\citep{Gaia:2022vcs,Delchambre:2022ugo} estimates redshifts of quasars from their BP/RP spectra. All sources having reliable~\citep{Delchambre:2022ugo} estimated redshift also enter the list.    

    \item All sources classified as AGN by the Vari module are included. This module uses photometric light curves to characterise variability.

    \item The Extended Objects (EO) module analyses surface brightness profiles of sources to look for physical extensions. Quasars analysed by this module are also included.

    \item All sources used to define the Gaia CRF3 (Celestial Reference Frame 3)~\citep{Gaia:2022huk} are included. These sources are cross-matched between Gaia and several external quasar catalogs and selected according to specific quality metrics.
\end{itemize}

However, this sample, with 6,649,162 sources, favors completeness rather than purity, so it very likely includes a lot of sources that are not actual QSOs. Therefore, we must reduce this sample to work with a cleaner dataset.

\subsection{Masked dataset}
\label{ssec:masked}

\begin{figure*}
	\centering
	\includegraphics[width=0.42\linewidth]{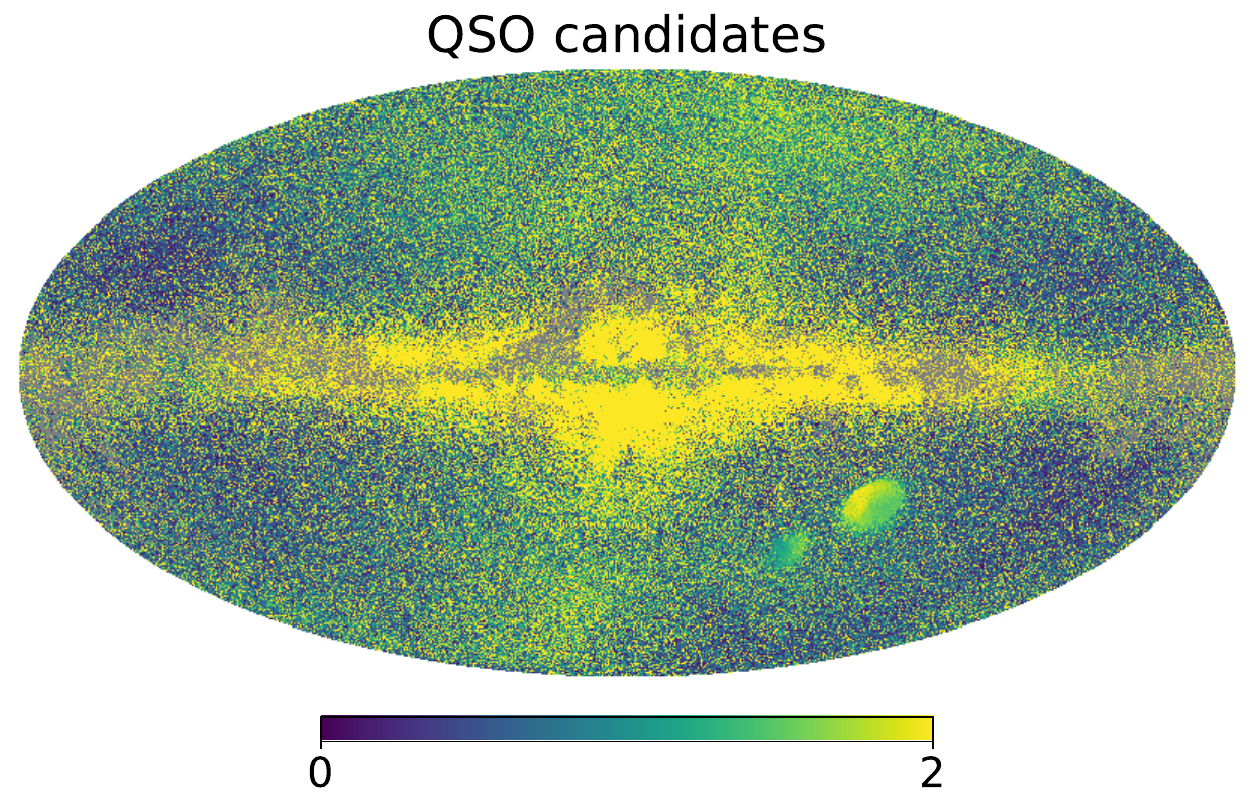}\includegraphics[width=0.42\linewidth]{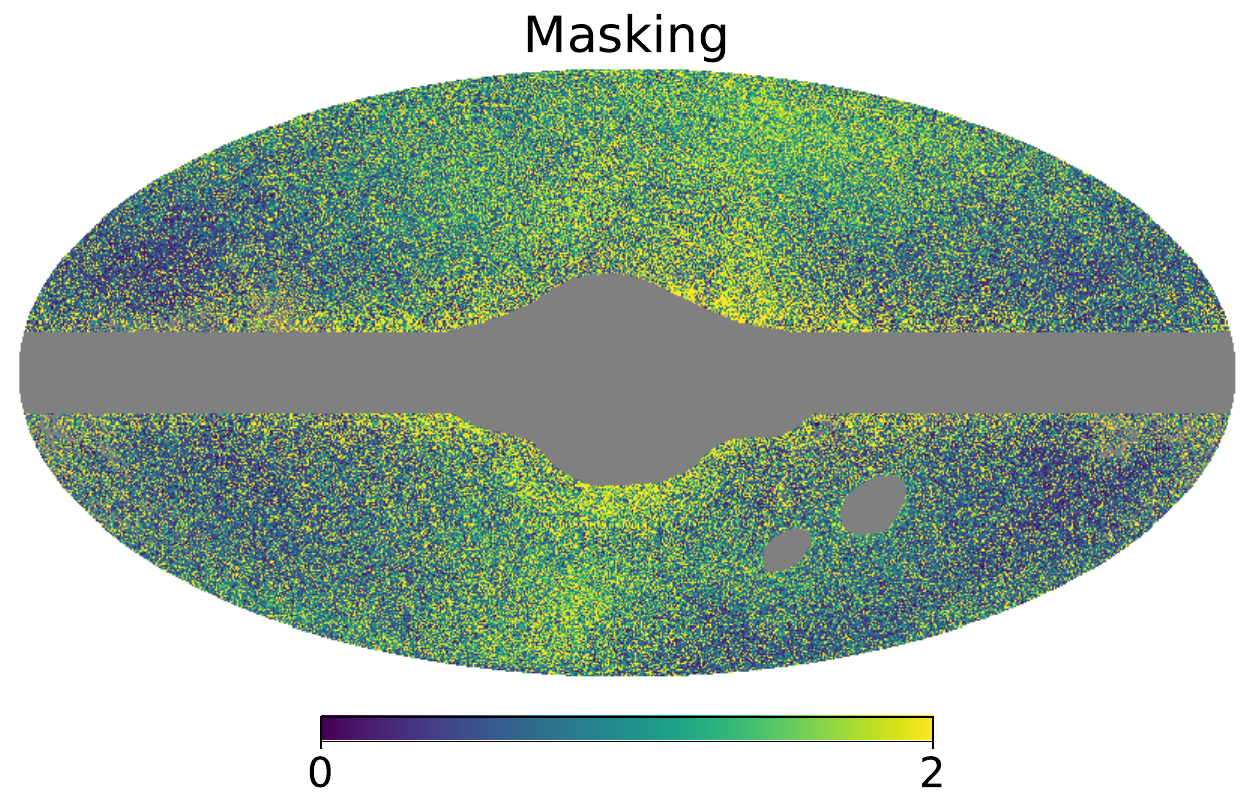}
	\includegraphics[width=0.42\linewidth]{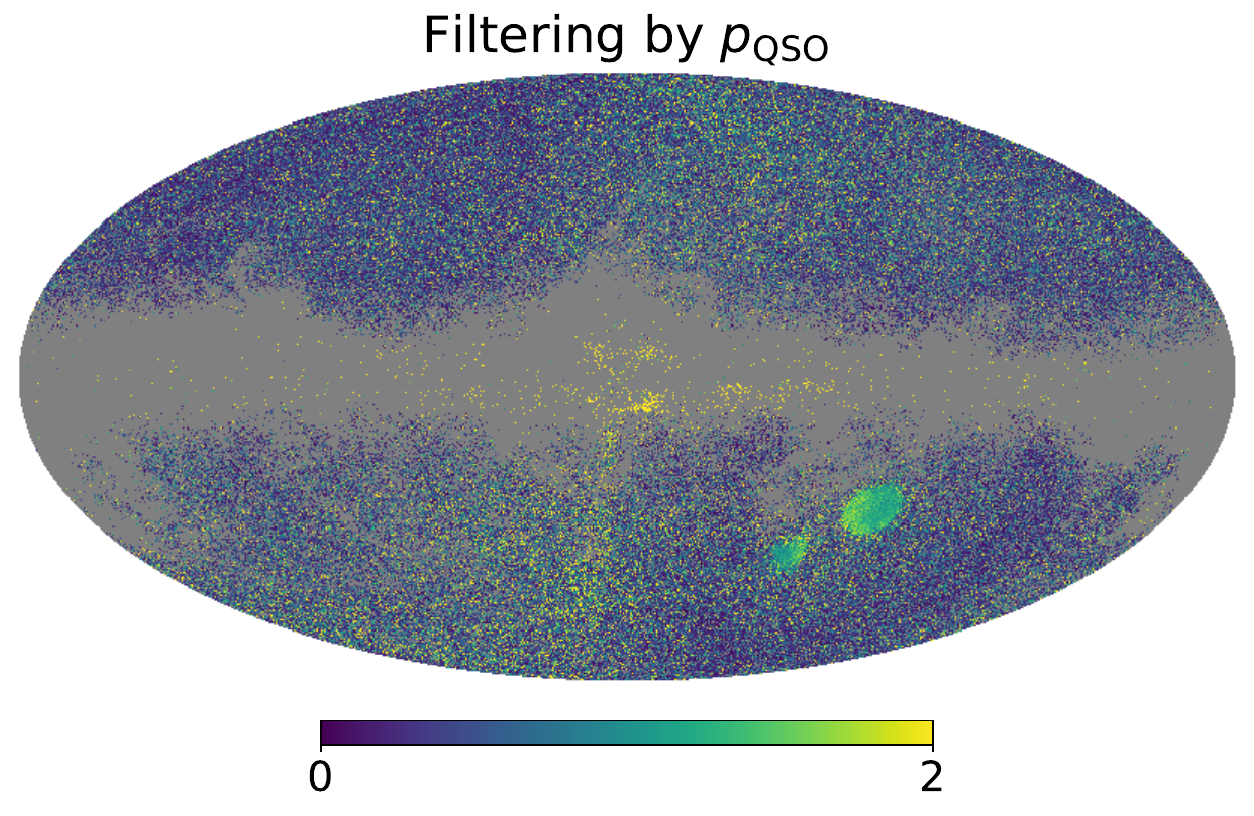}\includegraphics[width=0.42\linewidth]{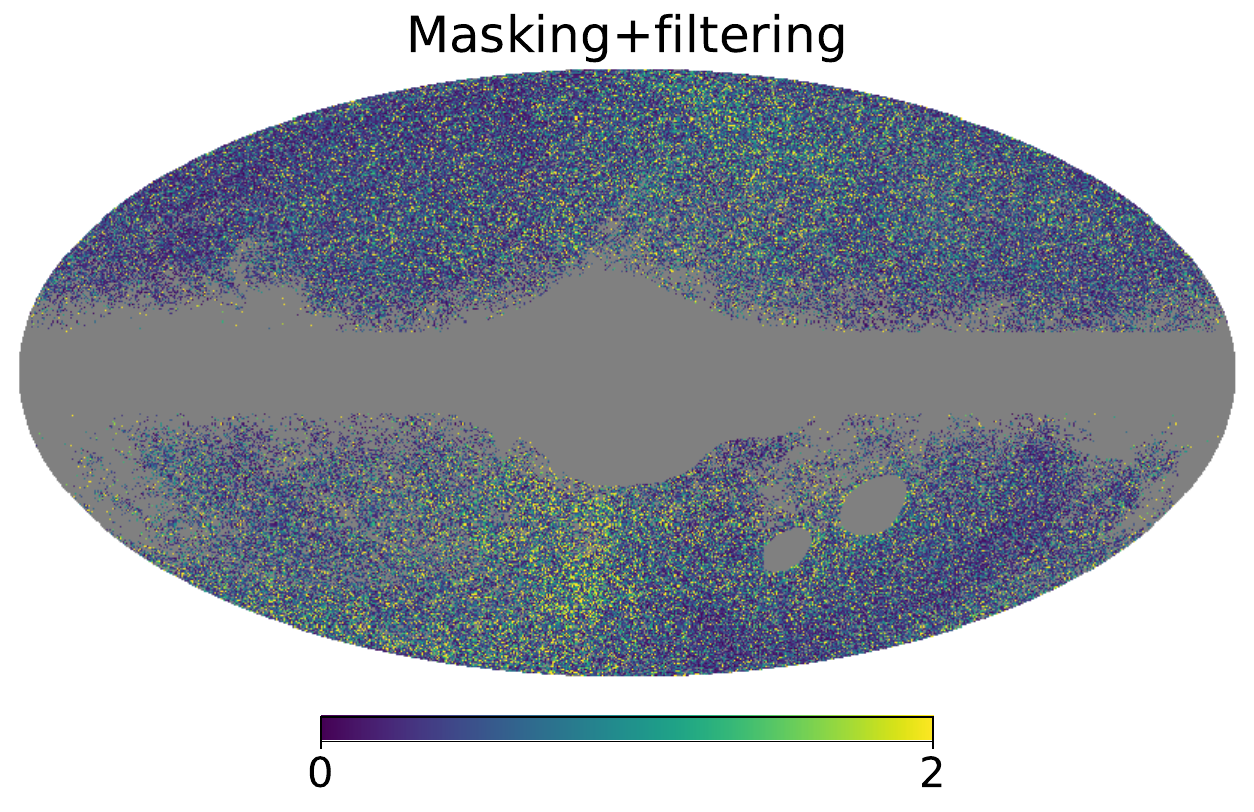}
	\caption{Proper motion module (in $\mathrm{mas/yr}$) skymap for different steps in our cleaning procedure, using \textit{HEALPix} level 8. The top left panel shows the full QSO candidates dataset, while the others show the resulting skymap after either of the steps or both.}
	\label{fig:pmskymaps_masks}

    \bigskip
    
    \includegraphics[width=0.42\linewidth]{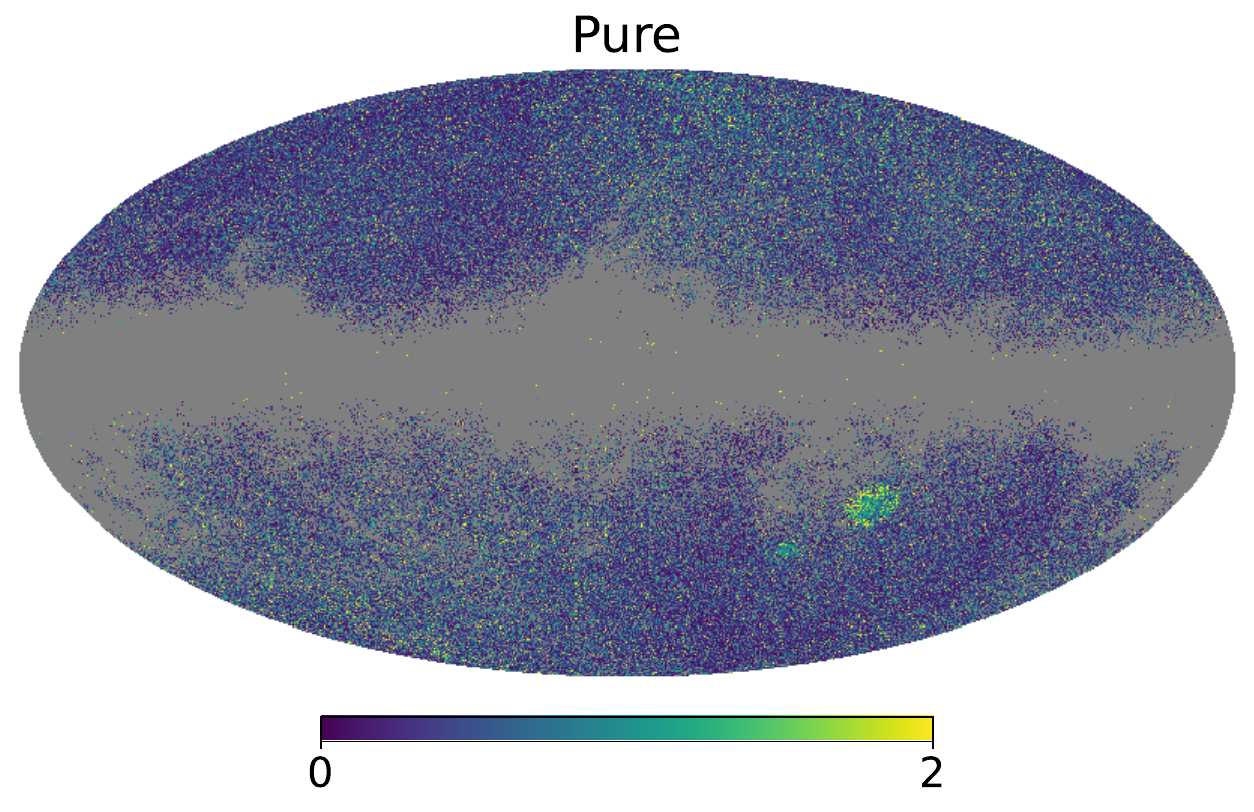}\includegraphics[width=0.42\linewidth]{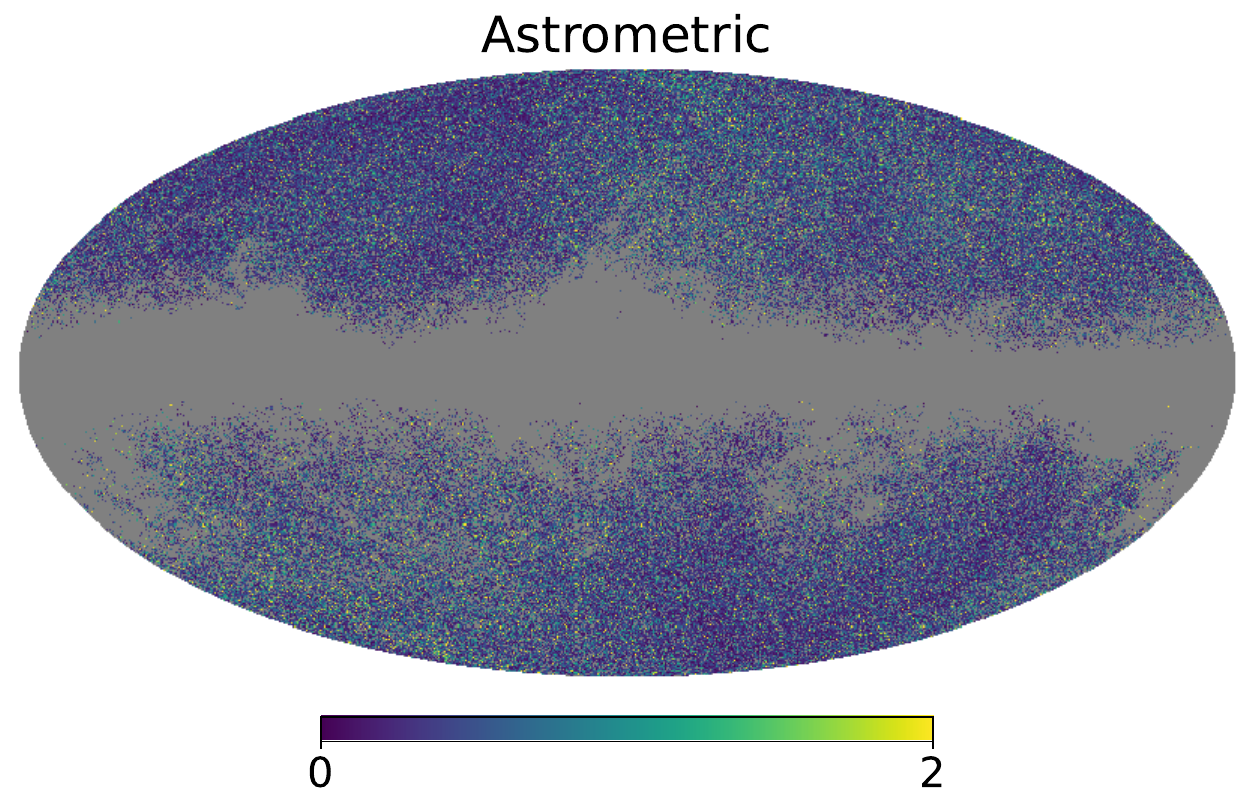}
    \includegraphics[width=0.42\linewidth]{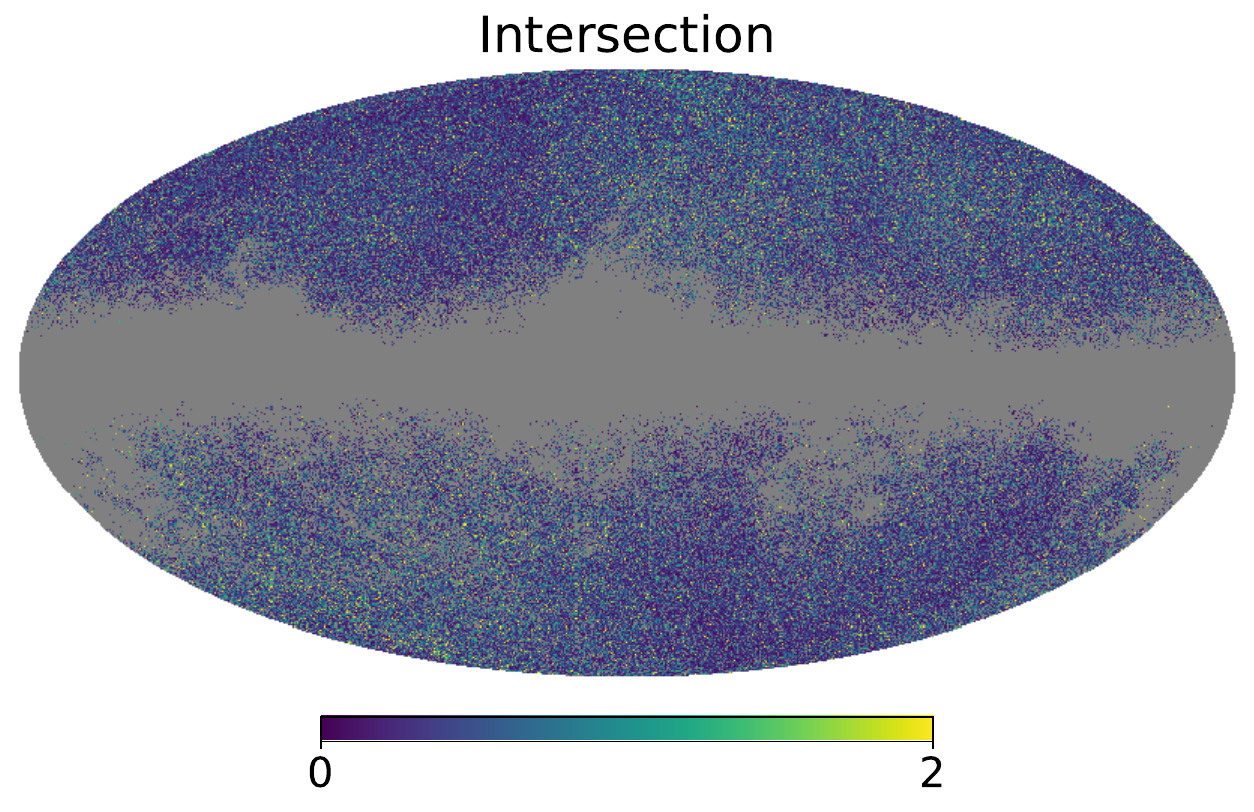}\includegraphics[width=0.42\linewidth]{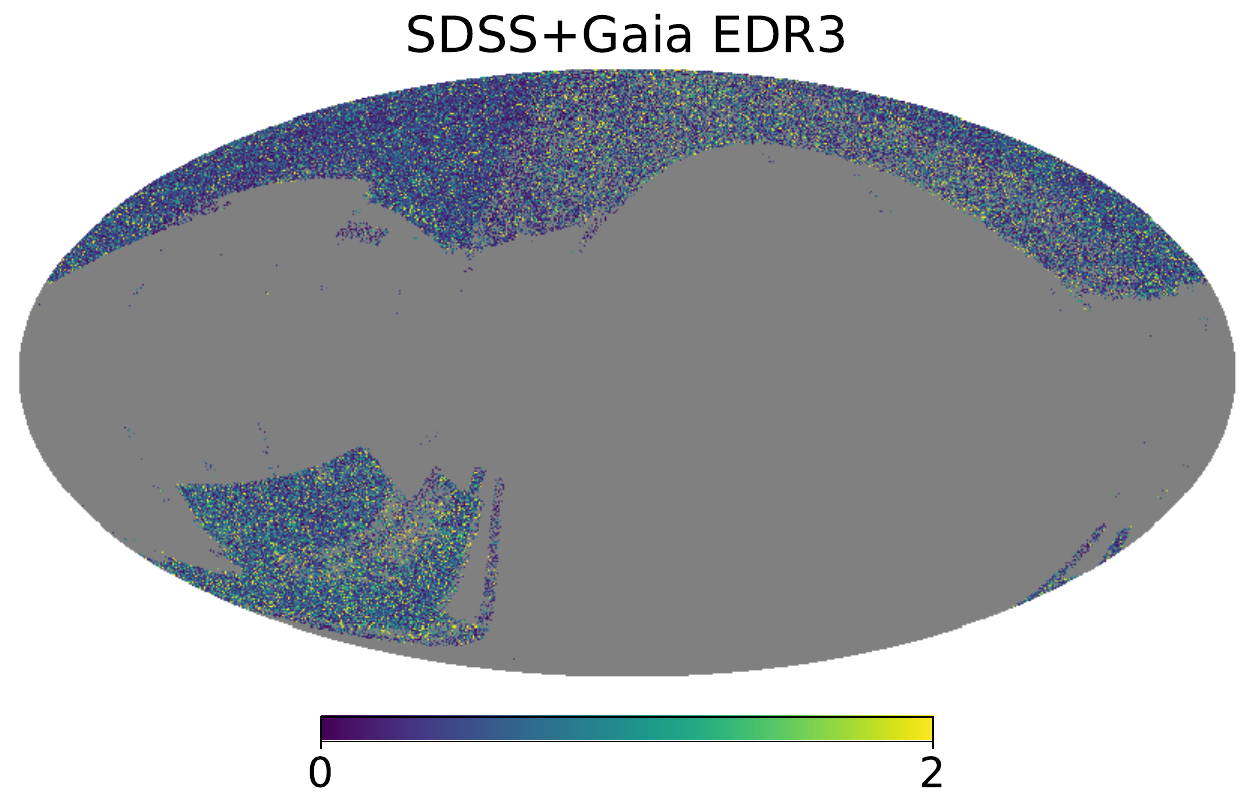}
	\caption{Proper motion module (in $\mathrm{mas/yr}$) skymap for different considered datasets, using \textit{HEALPix} level 8.}
	\label{fig:pmskymaps}
\end{figure*}

As we can see in the top left panel in Fig.~\ref{fig:pmskymaps_masks}, the original Gaia DR3 QSO candidates dataset is heavily biased by objects with large proper motions in the Galactic plane and Magellanic clouds.

In order to clean this dataset, one thing we could do is setting a threshold in $p_{\rm QSO}$. However, this does not completely remove the high proper motion regions. Therefore, and given the fact that the Galactic plane and Magellanic clouds are the main contaminating regions, we decided to mask these areas.

Our cleaning process then has two steps:

\begin{itemize}
    \item First, we mask the Galactic plane and Magellanic regions. In the top right panel in Fig.~\ref{fig:pmskymaps_masks}, we show the original dataset with the mask, which reduces the sources to 3,240,636. The figures were generated with \textit{HEALPix} implementation in Python \textit{healpy}~\citep{Zonca:2019vzt, Gorski:2004by}.

    \item Then, we set a threshold in $p_{\rm QSO}$, which we will generically refer to as \textit{filtering}. Setting a threshold very close to 1 gets rid of non-QSO sources at the cost of removing a lot of possibly good sources, drastically reducing the dataset size, so we need to find a good balance. Since we have a certain probability of being a quasar $p_{\rm QSO, i}$ for each source $i$, by assuming that all sources are independent, we can estimate the number of non-QSO sources in a dataset as the sum of expected values of each Bernoulli distribution, $\sum_i (1-p_{\rm QSO, i})$. We then set the threshold to the less constraining $p_{\rm QSO}$ which reduces this expected value to less than 1.

    In the bottom left panel of Fig.~\ref{fig:pmskymaps_masks}, we show the result of applying this last step to the original dataset, which reduces it to 1,154,431 sources, with a threshold in $p_{\rm QSO}$ of 0.99999465. As we can see, there is still notorious contamination from the Galactic plane and Magellanic clouds.
\end{itemize}

In the bottom right panel, we show the result after the two-step procedure, which is much cleaner and consists of 871,441 sources, with a threshold in $p_{\rm QSO}$ of 0.99998701. Note that this threshold is much less stringent than the previous one due to the preceding masking.

\subsection{Gaia QSO selections}
\label{ssec:gaia_qso}

In addition to our masked dataset, we consider other selections suggested in~\cite{Gaia:2022vcs}. This article provides several ways to get higher purity subsamples, of which the ones we use here are the following ones:

(i) Astrometric selection: sources are accepted or discarded based on astrometry criteria. All the sources in this selection must pass two filters:
	\begin{itemize}
		\item First step: individual sources with high-quality astrometric solutions and statistically insignificant parallaxes and proper motions are selected. More than 200 million sources match this criteria, mostly stars from our galaxy~\citep{Gaia:2022huk}.
		
		\item Second step: samples of sources with near-Gaussian distributions in uncertainty-normalized parallaxes and proper motions are selected.
	\end{itemize}
	
(ii) Pure selection. This refines the criteria used to construct the QSO candidates list:
	\begin{itemize}
		\item All sources from Gaia CRF3 are included.
		
		\item Sources from EO (Extended Objects) are included except for those with close neighbours.
		
		The previous classifiers use lists of quasars identified by other surveys, so their samples are believed to be above 90\% pure. On the other hand, the next two classifiers use supervised machine learning to discover new objects just using Gaia data.
		
		\item From the DSC sample, sources must be assigned the joint label of \textit{quasar} by the DSC, which requires that both Specmod and Allosmod assign probabilities to be above 0.5 of being a quasar. This subset is believed to have a purity of 62\%, increasing to 79\% when the Galactic plane ($|b|<11.54$ deg) is avoided.
		
		\item Sources from the Vari sample are included, given that they are considered to have a purity over 90\%. These results already exclude the Galactic plane.
	\end{itemize}

The astrometric selection has 1,897,754 sources and around 98\% purity or better~\citep{Gaia:2022vcs}. The pure one has 1,942,825 sources and a purity of 96\%. Their intersection, which is also considered as a separate dataset, has 1,801,255 sources.

Given the high number of sources and from the $p_{\rm QSO}$ distribution of these datasets, we conclude there is still significant contamination from non-QSO sources (of order 400,000). Therefore, we apply the same filtering procedure as for the masked dataset, setting a threshold in $p_{\rm QSO}$ to reduce the expected number of non-QSO sources to one. Details on the thresholds and final numbers of sources for each dataset are provided in Table~\ref{tab:redshifts}, while skymaps are displayed in Fig.~\ref{fig:pmskymaps}.

We can see that the three selections do a good job removing contamination from our galaxy, but the pure one does not completely remove the contamination from the Magellanic clouds. We do the fits for these three datasets as well as the masked one. However, it is clear that the cleaner dataset we have is the intersection between the astrometric and pure selections, followed by the astrometric one. Therefore, the latter are the two results which should be considered more rigorous.

\subsection{Datasets from previous works}
\label{ssec:previous_datasets}

Given the similarities of our analysis with the ones done by~\cite{Darling:2018hmc} and~\cite{Aoyama:2021xhj}, it is worth considering the same datasets they used and checking the consistency of our methodology. For this purpose, we consider three additional datasets.

\cite{Darling:2018hmc} use the VLBA catalog in~\cite{Truebenbach:2017nhp} excluding two sources with very high proper motion, which leaves a dataset with 711 sources. We consider the same dataset with the proper motions published in~\cite{Truebenbach:2017nhp}, which may differ slightly from the ones used in~\cite{Darling:2018hmc} given that they obtain the proper motions from bootstrap-resampled time series, but the datasets should be statistically consistent.

In addition, for some of these sources, Gaia DR1 provides additional data. The authors then use this information to extend the time series for these sources and generate a new catalog of 577 sources, of which they finally use 508. We refer to this sample as the VLBA+Gaia DR1 dataset.

Finally, \cite{Aoyama:2021xhj} cross-correlate the 16th data release of the Sloan Digital Sky Survey (SDSS) QSO catalog~\citep{Lyke:2020tag} with the astrometric data in Gaia EDR3. We use the SDSS+Gaia EDR3 cross-matched data provided by the Gaia collaboration\footnote{Gaia EDR3 documentation: \url{https://gea.esac.esa.int/archive/documentation/GEDR3/Catalogue_consolidation/chap_crossmatch/sec_crossmatch_externalCat/ssec_crossmatch_sdss.html}}, although we have not been able to get the exact number of sources as in~\cite{Aoyama:2021xhj}, obtaining 401,735. We do not apply the same filtering procedure as in previous datasets in order to keep this one comparable to the dataset in~\cite{Aoyama:2021xhj}. However, we have checked the SDSS+Gaia EDR3 filtered dataset to provide similar results as without the filter. A skymap of the SDSS+Gaia EDR3 dataset is provided in the bottom-right panel of Fig.~\ref{fig:pmskymaps}.

\subsection{Redshifts and frequency range validity}
\label{ssec:redshifts}

The main four datasets we use in this paper are obtained from Gaia DR3, which spans a time interval of 1038 days (2.84 years). Thus, as explained in Sec.~\ref{ssec:frequency_range}, the high frequency bound is $f\lesssim 1/T=1.12\times 10^{-8}~\mathrm{Hz}$. For the low frequency bound, we impose the constraint that the GW wavelength is lower than the distance to 75\% of the sources by following \cite{Darling:2018hmc}.

As has been mentioned previously, the Gaia QSOC module estimates redshifts of quasars from their BP/RP spectra. It is applied only for sources with $p_{\rm QSO}>0.01$ and, even in this case, redshifts are not always available. Therefore, not all the sources in our datasets have a redshift determination. In any case, we assume that the redshift distribution within our dataset is similar to the distribution within the subsample with determined redshifts.

In Table~\ref{tab:redshifts}, we show the number of sources with available redshift for each dataset together with the 25th percentile redshift and the corresponding distance and frequency, along with other data for reference. The distance is calculated assuming the cosmological parameters as $H_0=70$ km/s/Mpc, $\Omega_m=0.31$, and $\Omega_\Lambda=0.69$.

\begin{table*}
	\begin{tabular}{|c|c|c|c|c|c|c|c|c|}
		\hline
		Dataset & $N$ & $N_z$ & $z_{25}$ & $z_{50}$ & $z_{75}$ & $t_{25}$ (Gyr) & $f_{25}~(10^{-18}~\mathrm{Hz})$ & $(1-p_{\rm QSO}^{\rm min})\times 10^5$ \\\hline
		Masked & 871,441 & 871,438 & 0.961 & 1.629 & 2.408 & 7.52 & 4.21 & $1.299$ \\
        Pure & 816,641 & 816,640 & 0.967 & 1.621 & 2.374 & 7.54 & 4.20 & $1.924$ \\
		Astrometric & 786,165 & 786,164 & 0.977 & 1.622 & 2.359 & 7.58 & 4.18 & $1.959$ \\
		Intersection & 773,471 & 773,470 & 0.973 & 1.615 & 2.351 & 7.57 & 4.19 & $2.043$ \\\hline
        VLBA & 711 & 657 & 0.57 & 1.08 & 1.63 & 5.50 & 5.76 & - \\
        VLBA+Gaia DR1 & 508 & 483 & 0.63 & 1.12 & 1.64 & 5.87 & 5.40 & - \\
        SDSS+Gaia EDR3 & 401,735 & 392,993 & 1.032 & 1.764 & 2.543 & 7.80 & 4.06 & - \\\hline
	\end{tabular}
	\caption{The number of sources in each considered dataset, along with the number of sources with redshift, the 25th, 50th and 75th percentiles, the time of arrival to the source in the 25th percentile and its corresponding frequency. We also provide the thresholds in $p_{\rm QSO}$ for our four main datasets.}
	\label{tab:redshifts}
\end{table*}

As we can see in Table~\ref{tab:redshifts}, the derived frequencies are similar within 0.5\%. Therefore, we can roughly set this lower frequency to $4.2\times 10^{-18}~\mathrm{Hz}$. In summary, our derived upper bound for the SGWB is valid in the frequency range $4.2\times 10^{-18}~\mathrm{Hz}\lesssim f\lesssim 1.1\times 10^{-8}~\mathrm{Hz}$.

\section{Results}
\label{sec:results}

The main results of this paper are summarized in Table~\ref{tab:results}, where we list the fitted total power of the quadrupole moment, Z scores for $l=2$ ($Z_2$), Bayes factors $\mathcal{B}^{12}_1$ between the dipole+quadrupole and only dipole hypotheses, the best-fit $\Omega_{\rm GW}$ value corresponding to the quadrupole power, and the 95\% upper bound on $\Omega_{\rm GW}$.

\begin{table*}
	\begin{tabular}{|c|c|c|c|c|c|}
		\hline
		Dataset & $\sqrt{P_2}~(\mathrm{\mu as/yr})$ & $Z_2$ & $\ln\mathcal{B}^{12}_{1}$ & $h_{70}^2\Omega_{\rm GW}$ & $h_{70}^2\Omega_{\rm GW}^{\rm up}~(95\%)$ \\\hline
        Masked & 12.51(1.81) & 4.19 & -17.2 & 0.069(0.021) & 0.114 \\
        Pure & 23.15(2.01) & 10.21 & 34.4 & 0.235(0.040) & 0.295 \\
		Astrometric & 10.13(1.73) & 3.10 & -23.2 & 0.045(0.017) & 0.089 \\
		Intersection & 9.53(1.73) & 2.68 & -23.5 & 0.040(0.017) & 0.087 \\\hline
        VLBA & 2.73(1.23) & -1.93 & -42.3 & 0.0033(0.0056) & 0.024 \\
		VLBA+Gaia DR1 & 5.30(1.36) & 0.57 & -14.7 & 0.0123(0.0077) & 0.034 \\
		SDSS+Gaia EDR3 & 52.48(10.88) & 4.70 & 69.6 & 1.21(0.54) & 2.43
		 \\\hline
	\end{tabular}
	\caption{The total quadrupole power obtained from the dipole+quadrupole fits, together with the Z score corresponding to the quadrupole, the Bayes factor between the dipole+quadrupole and only dipole hypotheses, and the $\Omega_{\rm GW}$ estimations (best-fit and 95\% CL upper limit values). The values correspond to the maximum likelihood estimates and the 1-sigma errors are provided in brackets.}
	\label{tab:results}
\end{table*}

\begin{figure}
    \centering
    \includegraphics[width=\linewidth]{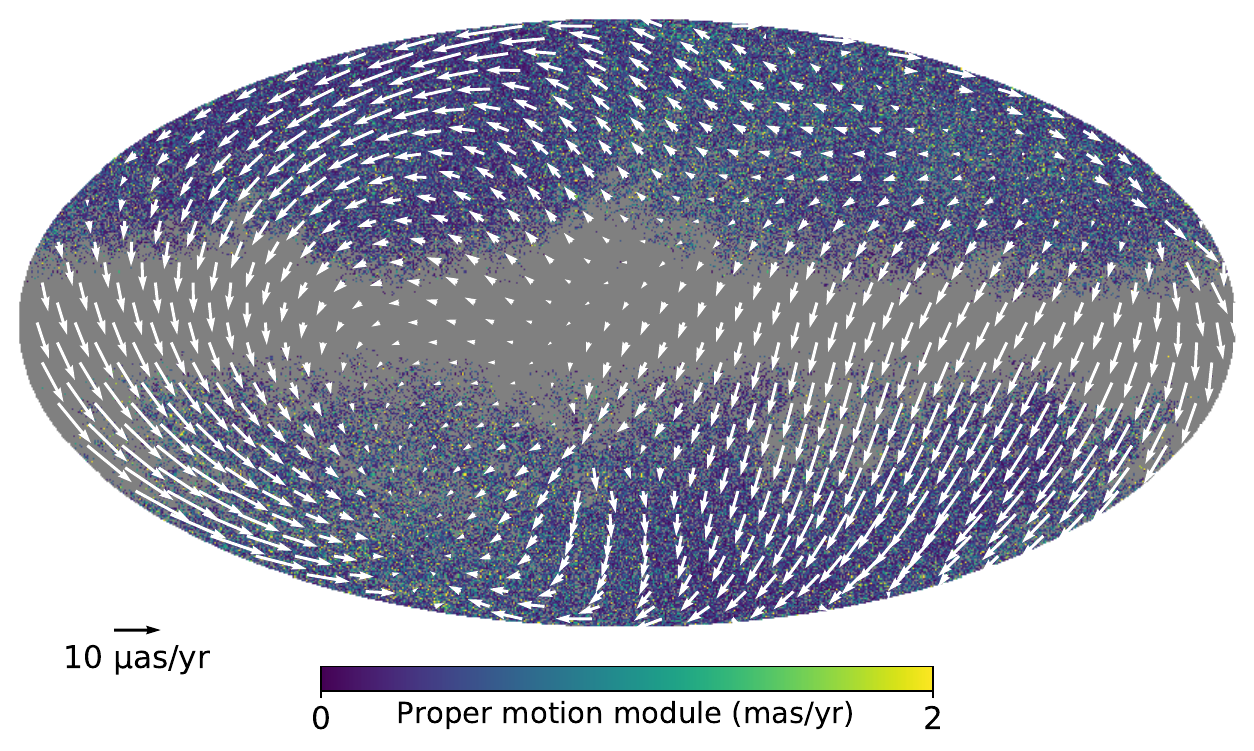}
    \caption{Fitted vector field skymap for the intersection dataset. The arrow scale is only orientative, given the distortion by the Mollweide projection.}
    \label{fig:vector_field}
\end{figure}

In addition, we also provide the fitted multipole coefficients for the four main datasets we consider in Table~\ref{tab:multipoles}, as well as the vector field for the intersection dataset in Fig.~\ref{fig:vector_field}. We refer the reader to App.~A for triangle plots showing the posterior distribution of the multipole coefficients and derived posteriors of $\Omega_{\rm GW}$ for all datasets. In Appendix~A we also show the full corner plots and correlation matrices for all runs.

\begin{table}
    \begin{tabular}{|c|c|c|c|c|}
        \hline
        & Masked & Pure & Astrometric & Intersection \\\hline
        $s_{10}$ & -39.92(1.92) & -4.40(1.89) & -5.72(1.89) & -4.79(1.88) \\
        $s_{11}^{\rm Re}$ & 11.92(1.49) & 4.64(1.44) & 0.31(1.46) & 0.18(1.44) \\
        $s_{11}^{\rm Im}$ & 3.74(1.45) & -6.02(1.42) & -8.86(1.43) & -9.06(1.42) \\
        $t_{10}$ & -14.01(2.09) & -5.81(2.04) & -0.31(2.04) & -0.20(2.06) \\
        $t_{11}^{\rm Re}$ & -14.19(1.44) & -6.48(1.39) & -1.57(1.38) & -1.60(1.41) \\
        $t_{11}^{\rm Im}$ & 5.91(1.33) & 9.96(1.29) & 3.74(1.30) & 3.41(1.31) \\\hline
        $\sqrt{P_1^s}$ & 43.65(1.91) & 11.60(1.70) & 13.78(1.88) & 13.68(1.74) \\
        $\sqrt{P_1^t}$ & 25.86(2.12) & 17.78(1.96) & 5.75(1.80) & 5.32(1.82) \\\hline
        $\sqrt{P_1}$ & 50.74(2.13) & 21.23(1.99) & 14.93(1.95) & 14.68(1.94) \\\hline
        $s_{20}$ & 3.84(1.88) & -8.65(1.81) & -5.79(1.81) & -5.15(1.82) \\
        $s_{21}^{\rm Re}$ & -0.85(1.28) & -8.33(1.24) & 2.29(1.24) & 2.10(1.24) \\
        $s_{21}^{\rm Im}$ & 5.16(1.37) & -0.91(1.33) & 1.85(1.33) & 1.91(1.34) \\
        $s_{22}^{\rm Re}$ & 2.15(1.40) & 4.49(1.37) & -0.65(1.38) & -0.50(1.37) \\
        $s_{22}^{\rm Im}$ & 1.77(1.41) & 5.29(1.38) & 2.63(1.37) & 2.01(1.38) \\
        $t_{20}$ & 5.25(1.98) & -1.69(1.94) & 3.63(1.93) & 3.74(1.95) \\
        $t_{21}^{\rm Re}$ & -1.09(1.39) & 3.40(1.36) & -0.29(1.35) & 0.72(1.36) \\
        $t_{21}^{\rm Im}$ & 3.59(1.43) & -6.74(1.41) & 1.72(1.41) & 1.84(1.41) \\
        $t_{22}^{\rm Re}$ & -1.59(1.32) & -7.16(1.29) & -2.56(1.29) & -2.82(1.29) \\
        $t_{22}^{\rm Im}$ & -2.33(1.34) & -1.52(1.31) & 1.52(1.30) & 0.97(1.31) \\\hline
        $\sqrt{P_2^s}$ & 9.21(1.86) & 17.65(1.97) & 8.10(1.63) & 7.15(1.70) \\
        $\sqrt{P_2^t}$ & 8.46(2.02) & 14.97(2.25) & 6.08(1.80) & 6.28(2.01) \\\hline
        $\sqrt{P_2}$ & 12.51(1.81) & 23.15(2.01) & 10.13(1.73) & 9.53(1.73) \\\hline
    \end{tabular}
    \caption{Fitted multipole coefficients, in $\mathrm{\mu as/yr}$, for the four main considered datasets. The values correspond to the maximum likelihood estimates and the 1-sigma errors are provided in brackets.}
	\label{tab:multipoles}
\end{table}

For most of the datasets, we get moderate Z scores ($Z_2>2.5$), which, according to the discussion in Sec.~\ref{ssec:statistical_significance}, can be interpreted as sigma deviations. This means these datasets likely do not consist of pure isotropic Gaussian noise, which can be explained by possible biases in the datasets, like intrinsic proper motion contamination. This explains the highest Z score in the pure dataset, which is the one showing more contamination from the Galactic plane and Magellanic clouds.

The Bayes factors reinforce this point, since they do not detect any significant quadrupole contribution in general. The only exceptions are the pure dataset, whose quadrupole could be induced by the above-mentioned contamination, and the SDSS+Gaia EDR3, which has strong dipole-quadrupole correlations (see Fig.~\ref{fig:cov_matrices}), making the Bayes factor computation not reliable. In fact, comparing against a null signal hypothesis produces a negative log-Bayes factor in this particular case.

Among our four main datasets of Gaia DR3, we get similar upper bounds on $\Omega_{\rm GW}$. Considering that the intersection dataset is the cleanest one, we conclude that the upper bound on the SGWB amplitude from Gaia DR3 is $h_{70}^2\Omega_{\rm GW}\lesssim 0.087$ for $4.2\times 10^{-18}~\mathrm{Hz}\lesssim f\lesssim 1.1\times 10^{-8}~\mathrm{Hz}$. In any case, the highest discrepancy is obtained for the pure dataset, which has the above-mentioned caveats, but even in this case, the derived 95\% upper bound is within a factor four. Thus, our results are robust under different selections of datasets, and it is unlikely that one can derive a much more stringent constraint from Gaia DR3 data alone. In Table~\ref{tab:multipoles}, we can also see (relative) toroidal dipole powers comparable to the ones in the VLBA and VLBA+Gaia DR1 datasets in~\cite{Darling:2018hmc} ($P_1^t/P_1^s\lesssim 1/4$), which is a good indication of the purity of our data, except for the pure dataset.

We note that our Z score for VLBA and VLBA+Gaia DR1, as well as the constraints on the multipole coefficients and the derived bounds on $\Omega_{\rm GW}$, are slightly more pessimistic than those in~\cite{Darling:2018hmc}. The difference traces back to a different implementation of the likelihood~\eqref{eq:logl} in~\cite{Darling:2018hmc}, as well as the simultaneous fit to the dipole and quadrupole~\cite{Darling}. Note that, in this case, the lower frequency bound is $f\gtrsim 1.4\times 10^{-9}$ Hz, due to the observing period of 22.2 years, while the upper frequency bounds are provided in Table~\ref{tab:redshifts}. For the SDSS+Gaia EDR3 dataset, we are not able to reproduce the results in~\cite{Aoyama:2021xhj}. As seen in the bottom-right panel of Fig.~\ref{fig:pmskymaps}, this dataset covers a smaller portion of the sky, and the simultaneous dipole+quadrupole fits could be the reason of the worse constraints compared to the other datasets, as hinted by the posteriors with strong correlations seen in Figs.~\ref{fig:posteriors_SDSS} and~\ref{fig:cov_matrices}. If we only fit the quadrupole, as done in~\cite{Aoyama:2021xhj}, we get a better bound $h_{70}^2\Omega_{\rm GW}\lesssim 0.40$. However, there is still a large discrepancy with the results of \cite{Aoyama:2021xhj}. This could be because of the narrower prior range of the proper motion and different error estimation method applied in their paper.

Our derived constraint from Gaia DR3 data is a bit worse than the upper bounds obtained both from the VLBA and the VLBA+Gaia DR1 datasets, although they are within the same order of magnitude. The possible reason is the following. Using Eq.~\eqref{eq:omegagw_bound}, we can make a rough estimate for the upper bound on $\Omega_{\rm GW}$ for a mission having resolution of $\Delta\theta$. Since we are working with proper motion catalogs, we can rewrite it as
\begin{equation}
    \Omega_{\rm GW}\lesssim\frac{\Delta\mu^2}{NH_0^2},
\end{equation}
where we have assumed that the mean uncertainty in proper motion is roughly given by $\Delta\mu \sim \Delta\theta/T$. From this, we can see that, even though the Gaia DR3-based datasets have around $10^3$ times more sources, which should lead to a smaller quadrupole power, the higher uncertainties in proper motion ($\langle\Delta\mu\rangle\approx 670~\mathrm{\mu as/yr}$ versus $20~\mathrm{\mu as/yr}$) pushes in the opposite direction with a power of 2. Note, however, that this expression assumes that all the sources have the root mean square (rms) proper motion of order $\Delta\mu/\sqrt{N}$, which in practice is not true for real datasets. The results could also be affected by factors that are difficult to predict when we deal with real noise, which does not follow an ideal Gaussian and isotropic distribution, and by the unequal distribution of sources in the sky.

\section{Conclusions and future prospects}
\label{sec:conclusions}

Using Gaia DR3 datasets, we have obtained an upper bound for the SGWB of $h_{70}^2\Omega_{\rm GW}\lesssim 0.087$ for $4.2\times 10^{-18}~\mathrm{Hz}\lesssim f\lesssim 1.1\times 10^{-8}~\mathrm{Hz}$. As has been discussed, this is somewhat worse than the constraint set by the VLBA and the VLBA+Gaia DR1 datasets~\citep{Darling:2018hmc}, which we also reanalysed, mainly due to the larger errors in proper motion. This is because the mean time of observation in the VLBA dataset is 22.2 years, whereas Gaia DR3 spans an interval of just 2.84 years. Longer periods of observation have drastic effects for SGWB determination.

As future Gaia releases process longer time intervals, they will increase the signal-to-noise ratio (SNR) and reduce the uncertainties in proper motions. According to~\cite{Gaia:2022, Brown:2021}, Gaia data products increase the precision on the coordinate and parallax measurements by $\propto \sqrt{T}$ so, for Gaia DR4 (5.5 years) and DR5 (10 years), we expect an improvement factor of 1.4 and 1.9, respectively. On the other hand, the proper motion precision increases by $\propto T^{1.5}$, so the factors of improvement are 2.7 and 6.6 for DR4 and DR5, respectively. In Eq.~\eqref{eq:OmegaGW_pm}, we see that $\Omega_{\rm GW}$ is proportional to the proper motion squared. As a result, this leads to an improvement of $\propto T^3$, i.e. 7.2 and 44 for DR4 and DR5, respectively. If we take our current upper bounds of $\sim0.087$ and apply these improvement factors, the expected constraints are $\Omega_{\rm GW}\lesssim 0.012$ and $0.0020$ with the maximum constrained frequency going down to $5.8\times 10^{-9}~\mathrm{Hz}$ and $3.2\times 10^{-9}~\mathrm{Hz}$ for DR4 and DR5, respectively. This estimation assumes that our number of sources is going to remain constant over the time. However, with the increase in SNR, it is very likely that we will have a larger number of quasars in our dataset, which will further improve the constraint. In addition, with Gaia DR5, we will have access to each source full time series. As discussed in Sec.~\ref{ssec:frequency_range}, this will allow to constrain a different frequency region ($f\gtrsim 1/T\approx 3.2\times 10^{-9}~\mathrm{Hz}$).

In~\cite{Book:2010pf}, the Gaia sensitivity was estimated as $\Omega_{\rm GW}\lesssim 10^{-6}$ using $\Delta\theta=10~\mathrm{\mu as}$, $T=1~\mathrm{yr}$ and $N=10^6$. However, in~\cite{Darling:2018hmc}, it was argued that the final budget will be closer to an error in proper motion of $200~\mathrm{\mu as/yr}$ and $N=5\times 10^5$, which yields $\Omega_{\rm GW}\lesssim 10^{-3}$, closer to our estimate. The differences may be due to the assumption on the rms proper motion simply scaling with $\propto 1/\sqrt{N}$ as discussed in Sec.~\ref{sec:results}. \cite{Darling:2018hmc} also predicted $\Omega_{\rm GW}\lesssim 6\times10^{-4}$ from mock data based on a WISE-Gaia catalog~\citep{Paine:2018lav}. 

The proposed mission Theia will further improve the constraints set by Gaia. In~\cite{Garcia-Bellido:2021zgu}, a $T=20$~yrs Gaia mission is generously assumed, with angular resolution $\Delta\theta = 1$ mas and $N=10^9$ sources, producing a sensitivity of $\Omega_{\rm GW}\lesssim 10^{-8}$ at nHz frequencies. This estimate is very optimistic and leads to a similar prediction of $\Omega_{\rm GW}\lesssim 3\times10^{-14}$ for Theia, assuming a hundred times more objects and 60 times better angular resolution. However, if we multiply this factor of improvement ($60^2\times 100$) by the expected DR5 constraint $\Omega_{\rm GW}\lesssim 10^{-4}$ from the above discussion, we obtain that Theia will be able to set constraints of order $\Omega_{\rm GW}\lesssim 10^{-10}$. Such a sensitivity would be instrumental to constrain sources of stochastic backgrounds that are believed to emit at very low frequencies. Besides the guaranteed background from (unresolved) Super Massive Black Hole binaries~\citep{Sesana:2008mz,Burke-Spolaor:2018bvk}, other sources include cosmic strings~\citep{Blanco-Pillado:2017rnf,Matsui:2019obe}, phase transitions~\citep{Caprini:2010xv,Brandenburg:2021tmp} and signals from Primordial Black Holes~\citep{Saito:2008jc,Garcia-Bellido:2017aan,Braglia:2021wwa}. All these mechanisms were also proposed as interpretations of the common noise excess recently observed by the NANOGrav collaboration~\citep{NANOGrav:2020bcs}, and further timing of pulsars will probably have already either confirmed or ruled this possibility out by the time that Theia is accepted. Nevertheless, future astrometric surveys will offer a longer lever, which will help characterize such signals over a wider frequency range.

\section*{Acknowledgements}

We warmly thank Jeremy Darling and Shohei Aoyama for very useful correspondence on the datasets used in \cite{Truebenbach:2017nhp,Paine:2018lav,Darling:2018hmc} and \cite{Aoyama:2021xhj}, respectively, as well as Deyan Mihaylov for stimulating conversations about constraining gravitational waves with astrometry.
The authors acknowledge support from the Spanish Research Project PID2021-123012NB-C43 [MICINN-FEDER], and the Centro de Excelencia Severo Ochoa Program CEX2020-001007-S at IFT. S.J. is supported by the FPI grant PRE2019-088741 funded by MCIN/AEI/10.13039/501100011033. S.K. is supported by the Spanish Atracci\'on de Talento contract no. 2019-T1/TIC-13177 granted by Comunidad de Madrid, the I+D grant PID2020-118159GA-C42 funded by MCIN/AEI/10.13039/501100011033, the i-LINK 2021 grant LINKA20416 of CSIC, and Japan Society for the Promotion of Science (JSPS) KAKENHI Grant no. 20H01899 and 20H05853.  S.F. was supported by Physics and Astronomy Department of the University of Bologna through {\em Borsa di studio per la preparazione all'estero della tesi di Laurea Magistrale}. She thanks the IFT, where this work was carried out, for hospitality. This work has made use of data from the European Space Agency (ESA) mission {\it Gaia} (\url{https://www.cosmos.esa.int/gaia}), processed by the {\it Gaia} Data Processing and Analysis Consortium (DPAC, \url{https://www.cosmos.esa.int/web/gaia/dpac/consortium}). Funding for the DPAC has been provided by national institutions, in particular the institutions participating in the {\it Gaia} Multilateral Agreement. Some of the results in this paper have been derived using the healpy and HEALPix packages. The MCMCs have been run in the Hydra HPC cluster at the IFT.

\section*{Data Availability}

The main datasets used in this article can be obtained from the Gaia Archive with the queries indicated in~\cite{Gaia:2022vcs} and the thresholds for \texttt{classprob\_dsc\_combmod\_quasar} given in Table~\ref{tab:redshifts}. Any other data used or produced in this work is available upon reasonable request to the corresponding author.



\bibliographystyle{mnras}
\bibliography{main} 




\appendix

\section{Triangle plots from the MCMC analysis}
\label{app:triangle_plots}

\begin{figure*}
    \centering
    \includegraphics[width=\linewidth]{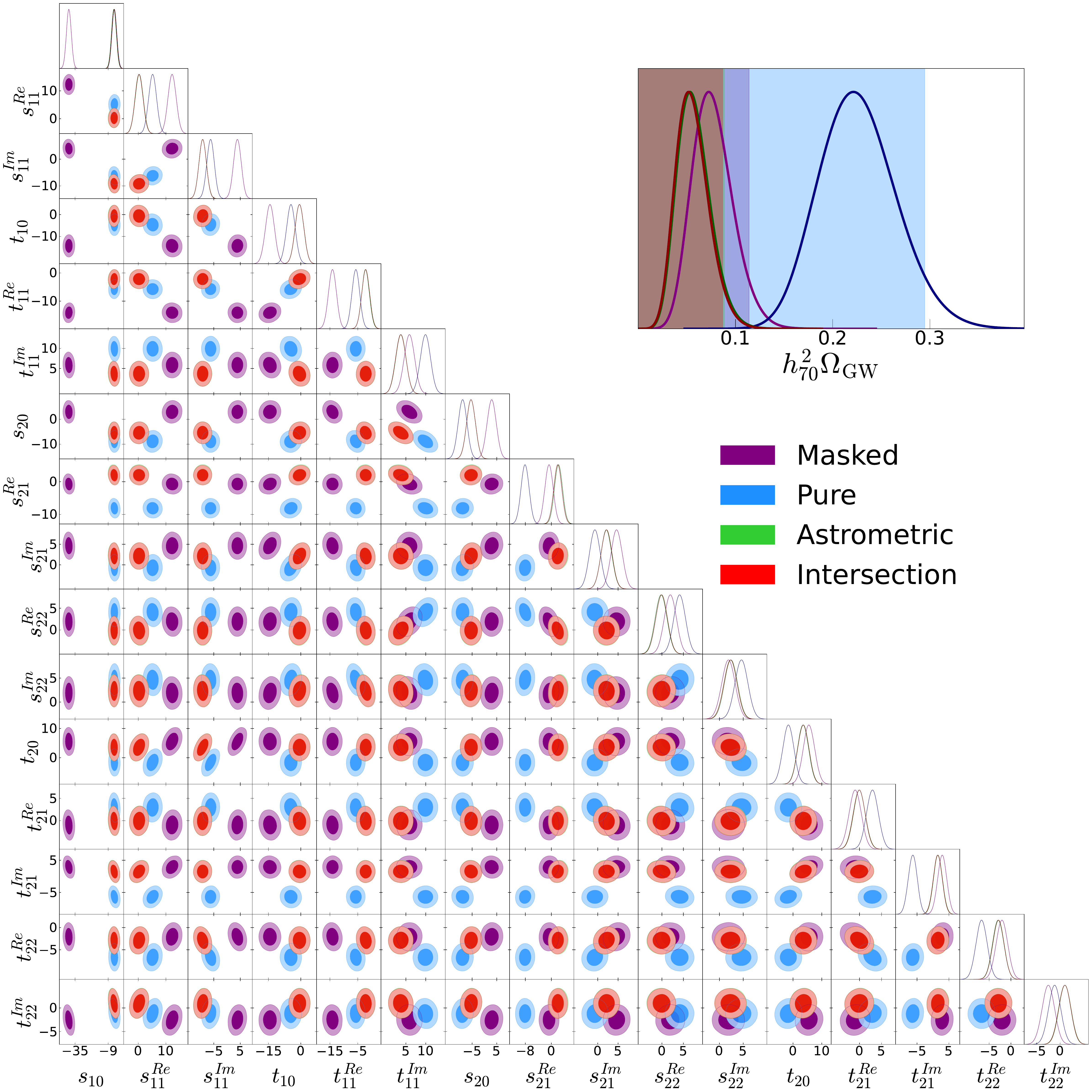}
    \caption{Posteriors for the masked, pure, astrometric and intersection datasets. Note that the results of the intersection and astrometric datasets considerably overlap.}
    \label{fig:posteriors_ours}
\end{figure*}

\begin{figure*}
    \centering
    \includegraphics[width=\linewidth]{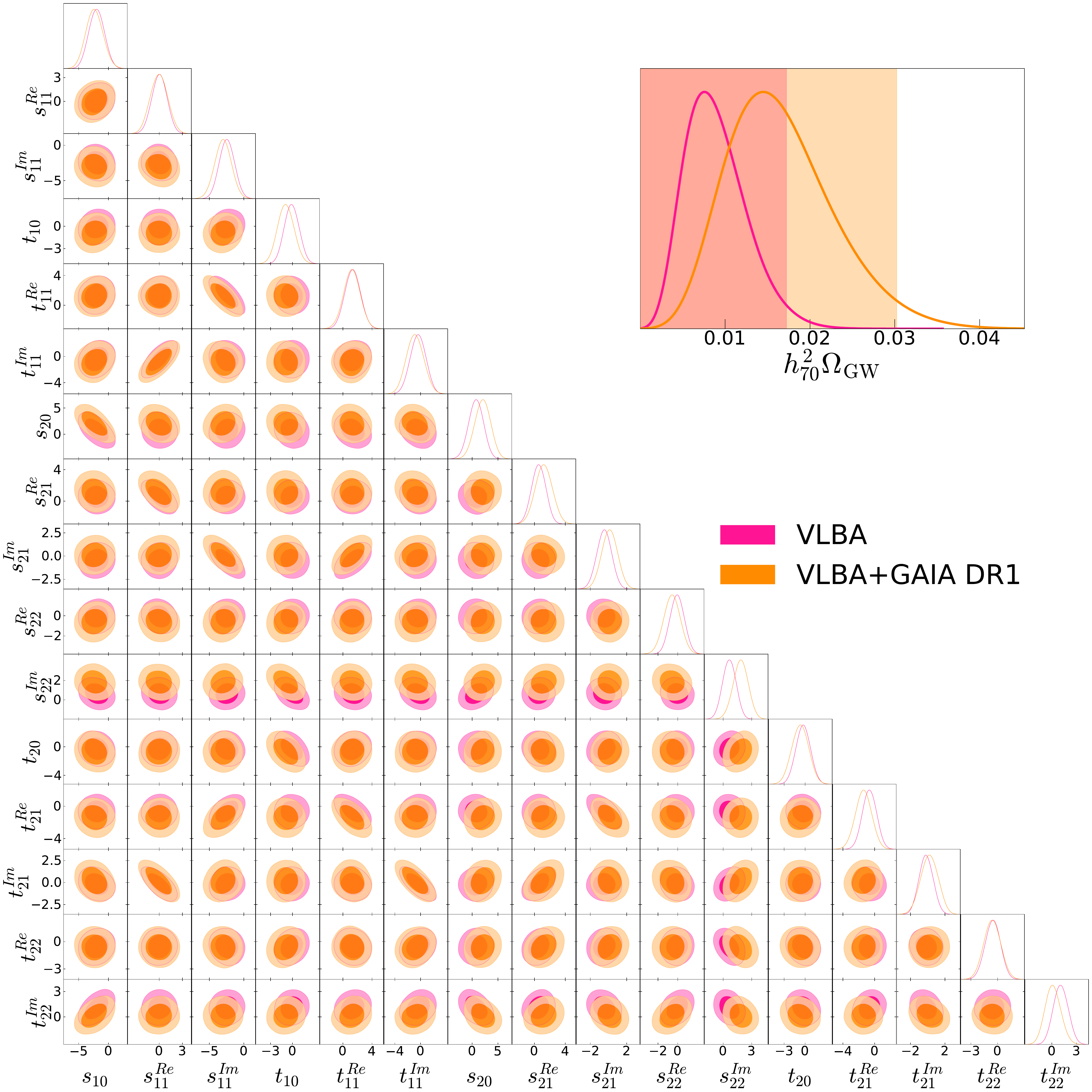}
    \caption{Posteriors for the VLBA and VLBA+Gaia DR1 datasets.}
    \label{fig:posteriors_VLBA}
\end{figure*}

\begin{figure*}
    \centering
    \includegraphics[width=\linewidth]{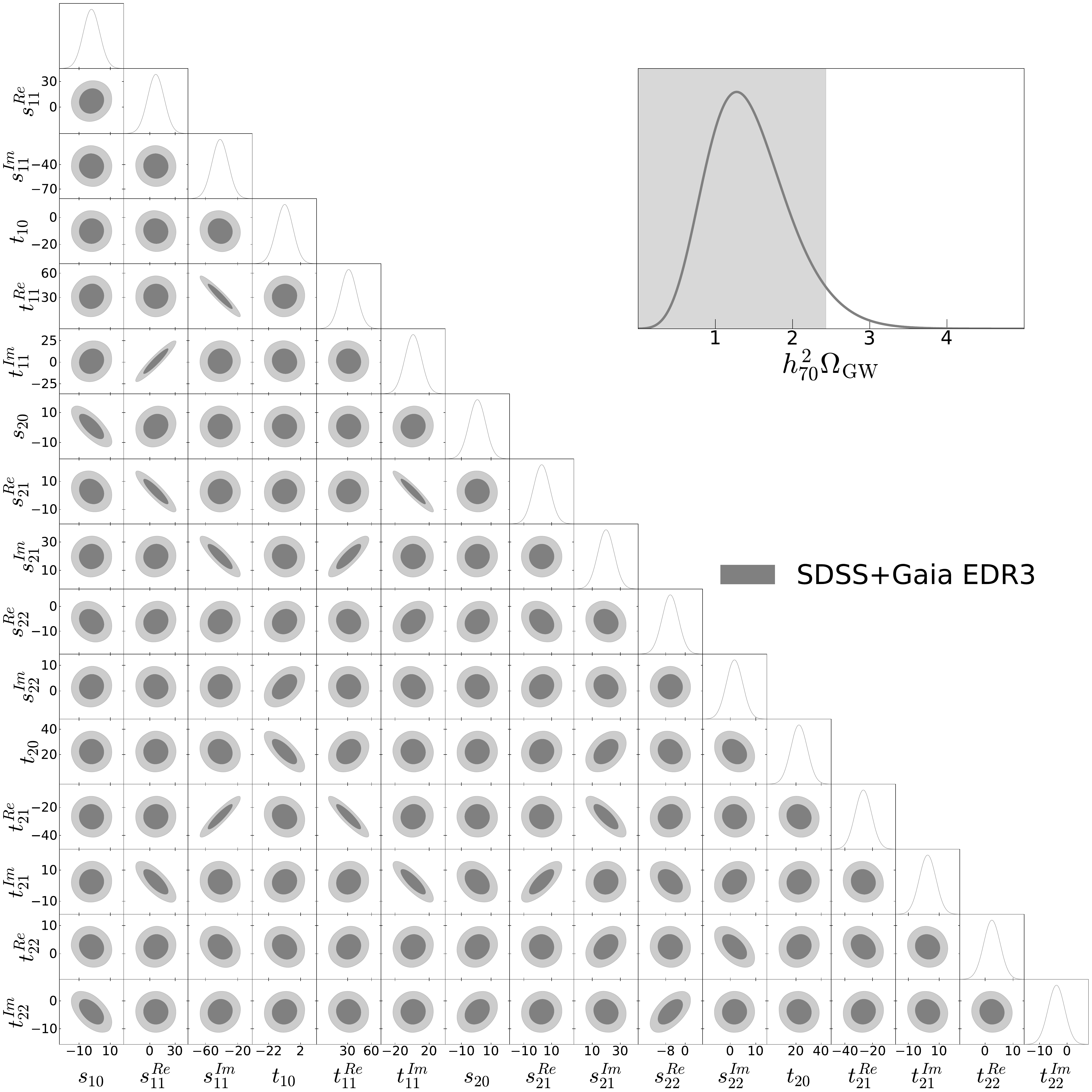}
    \caption{Posteriors for the SDSS+Gaia EDR3 dataset.}
    \label{fig:posteriors_SDSS}
\end{figure*}

\begin{figure*}
    \centering
    \includegraphics[width=\linewidth]{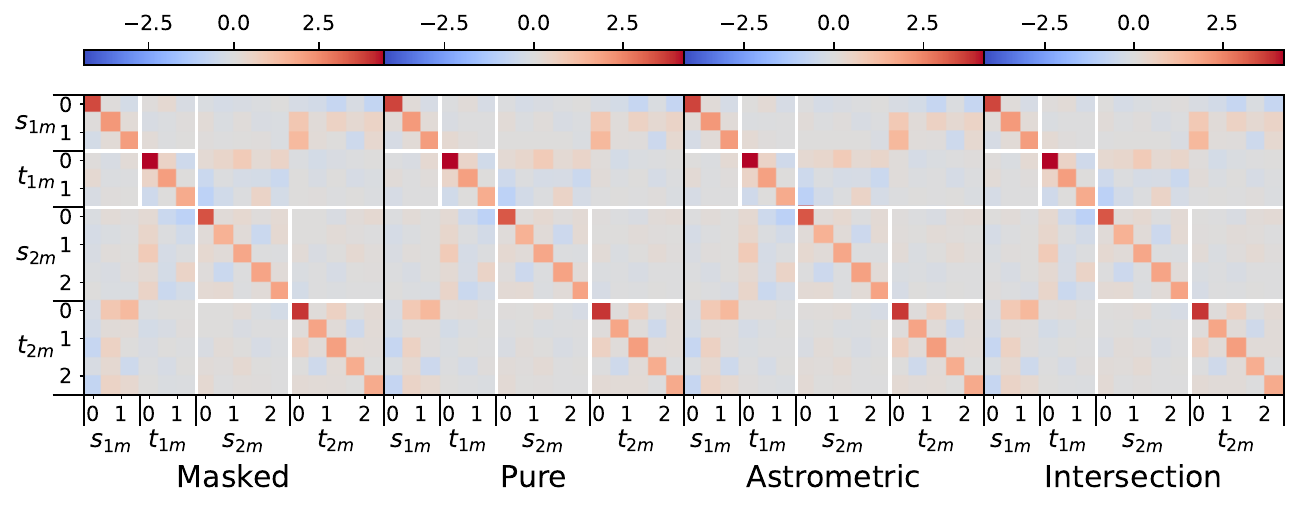}
    \includegraphics[width=0.7727\linewidth]{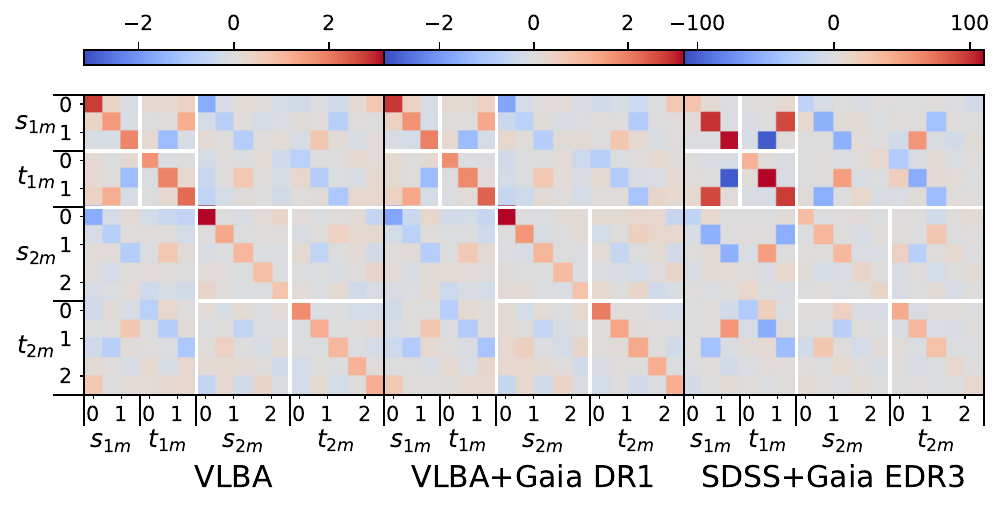}
    \caption{Covariance matrices for all the considered datasets, with units $(\mathrm{\mu as/yr})^2$.}
    \label{fig:cov_matrices}
\end{figure*}

In this Appendix, we supplement the constraints in the main text with triangle plots showing the marginalized 1 and 2-dimensional posterior distributions for the multipole coefficients $r_{\ell m}$ of the vector spherical harmonic decomposition of the proper motion field. We show triangle plots for all the datasets considered in our analysis in Figs.~\ref{fig:posteriors_ours}-\ref{fig:posteriors_SDSS}. In each figure, we also plot the derived posterior distribution for $\Omega_{\rm GW}$. Finally, in Fig.~\ref{fig:cov_matrices}, we show the covariance matrices from our analysis.


\bsp	
\label{lastpage}
\end{document}